\documentclass[sigconf,dvipsnames,table]{acmart}
\acmConference[ICSE '26]{IEEE/ACM 48th International Conference on Software Engineering}{April 12-18, 2026}{Rio de Janeiro, Brazil}

\author{
    Jiachi Chen\textsuperscript{1,2},
    Yiming Shen\textsuperscript{1},
    Jiashuo Zhang\textsuperscript{3*},
    Zihao Li\textsuperscript{4},
    John Grundy\textsuperscript{5},
    Zhenzhe Shao\textsuperscript{1}\\
    Yanlin Wang\textsuperscript{1},
    Jiashui Wang\textsuperscript{6},
    Ting Chen\textsuperscript{7,8},
    Zibin Zheng\textsuperscript{1}
}

\affiliation{%
 \textsuperscript{1}\textit{Sun Yat-sen University, Zhuhai, China}\\
 \textsuperscript{2}\textit{The State Key Laboratory of Blockchain and Data Security, Zhejiang University, Hangzhou, China}\\
\textsuperscript{3}\textit{Peking University, Beijing, China}; *Corresponding author\\
\textsuperscript{4}\textit{The Hong Kong Polytechnic University, Hong Kong, China}\\
\textsuperscript{5}\textit{Monash University, Melbourne, Australia}; 
\textsuperscript{6}\textit{Zhejiang University, Hangzhou, China}\\
\textsuperscript{7}\textit{University of Electronic Science and Technology of China, Chengdu, China}\\
\textsuperscript{8}\textit{Kashi Institute of Electronics and Information Industry, Kashi, China}\\
\country{}
}
\email{chenjch86@mail.sysu.edu.cn, shenym7@mail2.sysu.edu.cn, zhangjiashuo@pku.edu.cn}
\email{zi-hao.li@connect.polyu.hk, john.grundy@monash.edu, shaozhzh3@mail2.sysu.edu.cn}
\email{yanlin-wang@outlook.com, 12221251@zju.edu.cn, brokendragon@uestc.edu.cn, zhzibin@mail.sysu.edu.cn}

\AtBeginDocument{%
  \providecommand\BibTeX{{%
    \normalfont B\kern-0.5em{\scshape i\kern-0.25em b}\kern-0.8em\TeX}}}

\usepackage{caption}
\usepackage{tikz}
\usepackage{amsmath}
\usepackage{booktabs}
\usepackage{multirow}
\usepackage{makecell}
\usepackage{enumitem}
\usepackage{listings}
\usepackage[many]{tcolorbox}
\tcbuselibrary{listings}

\usepackage{algorithm}
\usepackage{algpseudocode}
\usepackage{appendix}
\usepackage{xspace}
\usepackage{lipsum}

\definecolor{customcite}{HTML}{b83b5e}
\definecolor{customlink}{HTML}{07689f}
\definecolor{customurl}{HTML}{11999e}

\AtEndPreamble{
    \usepackage{hyperref}
    \hypersetup{
      colorlinks = true,
      linkcolor = customlink,
      anchorcolor = purple,
      citecolor = customcite,
      filecolor = purple,
      urlcolor = customurl
    }

}

\definecolor{verylightgray}{rgb}{.97,.97,.97}
\lstdefinelanguage{Solidity}{
  keywords=[1]{anonymous, assembly, assert, balance, break, call, callcode, case, catch, class, constant, continue, constructor, contract, debugger, default, delegatecall, delete, do, else, emit, event, experimental, export, external, false, finally, for, function, gas, if, implements, import, in, indexed, instanceof, interface, internal, is, length, library, log0, log1, log2, log3, log4, memory, modifier, new, payable, pragma, private, protected, public, pure, push, require, return, returns, revert, selfdestruct, send, solidity, storage, struct, suicide, super, switch, then, this, throw, transfer, true, try, typeof, using, value, view, while, with, addmod, ecrecover, keccak256, mulmod, ripemd160, sha256, sha3}, 
  keywordstyle=[1]\color{blue}\bfseries,
  keywords=[2]{address, bool, byte, bytes, bytes1, bytes2, bytes3, bytes4, bytes5, bytes6, bytes7, bytes8, bytes9, bytes10, bytes11, bytes12, bytes13, bytes14, bytes15, bytes16, bytes17, bytes18, bytes19, bytes20, bytes21, bytes22, bytes23, bytes24, bytes25, bytes26, bytes27, bytes28, bytes29, bytes30, bytes31, bytes32, enum, int, int8, int16, int24, int32, int40, int48, int56, int64, int72, int80, int88, int96, int104, int112, int120, int128, int136, int144, int152, int160, int168, int176, int184, int192, int200, int208, int216, int224, int232, int240, int248, int256, mapping, string, uint, uint8, uint16, uint24, uint32, uint40, uint48, uint56, uint64, uint72, uint80, uint88, uint96, uint104, uint112, uint120, uint128, uint136, uint144, uint152, uint160, uint168, uint176, uint184, uint192, uint200, uint208, uint216, uint224, uint232, uint240, uint248, uint256, var, void, ether, finney, szabo, wei, days, hours, minutes, seconds, weeks, years},  
  keywordstyle=[2]\color{teal}\bfseries,
  keywords=[3]{block, blockhash, coinbase, difficulty, gaslimit, number, timestamp, msg, data, gas, sender, sig, value, now, tx, gasprice, origin},  
  keywordstyle=[3]\color{violet}\bfseries,
  identifierstyle=\color{black},
  sensitive=false,
  comment=[l]{//},
  morecomment=[s]{/*}{*/},
  commentstyle=\color{gray}\ttfamily,
  stringstyle=\color{red}\ttfamily,
  morestring=[b]',
  morestring=[b]"
}
\newtcolorbox{myboxc}{
    colback=gray!15!white,
    arc = 0pt, outer arc = 0pt,
    boxsep=0pt, left = 3pt, right = 0pt, top = 0pt, bottom = 0pt, 
    leftrule=3pt, bottomrule=0pt,toprule=0pt, rightrule=0pt,
    left = \boxmargin, right = \boxmargin, top = \boxmargin, bottom = \boxmargin
}
\lstdefinelanguage{JSON}{
    basicstyle=\ttfamily\small,
    numbers=left,
    numberstyle=\tiny,
    stepnumber=1,
    numbersep=8pt,
    showstringspaces=false,
    breaklines=true,
    stringstyle=\color{black},
    keywords=[1]{false, true, null, string, int, enum},
    keywordstyle=[1]\color{blue}\bfseries,
    morekeywords={},
    comment=[l]{//},
    commentstyle=\color{green!50!black},
    morecomment=[s]{/*}{*/},
    literate=
        *{0}{{{\color{blue}0}}}{1}
         {1}{{{\color{blue}1}}}{1}
         {2}{{{\color{blue}2}}}{1}
         {3}{{{\color{blue}3}}}{1}
         {4}{{{\color{blue}4}}}{1}
         {5}{{{\color{blue}5}}}{1}
         {6}{{{\color{blue}6}}}{1}
         {7}{{{\color{blue}7}}}{1}
         {8}{{{\color{blue}8}}}{1}
         {9}{{{\color{blue}9}}}{1}
         {:}{{{\color{black!80!white}{:}}}}{1}
         {,}{{{\color{black!80!white}{,}}}}{1}
         {\{}{{{\color{black!80!white}{\{}}}}{1}
         {\}}{{{\color{black!80!white}{\}}}}}{1}
         {[}{{{\color{black!80!white}{[}}}}{1}
         {]}{{{\color{black!80!white}{]}}}}{1},
}

\newtcolorbox{jsonbox}[1][]{
    listing only,
    colback=gray!5,
    colframe=gray!50!black,
    listing engine=listings,
    listing options={
        language=JSON,
        basicstyle=\ttfamily\small,
        breaklines=true,
        postbreak=\mbox{\textcolor{red}{$\hookrightarrow$}\space},
    },
    title=Structured Information,
    arc=1mm,
    left=5mm, right = 0mm,
    #1
}
\newtcolorbox{findingbox}{
    colback=gray!10,
    colframe=gray!50,
    boxrule=0.5pt,
    arc=1mm,
    boxsep=1.5mm,
    left=1.7mm,right=1.7mm,top=1mm,bottom=1mm,
    fontupper=\itshape
}
\newtcolorbox{notebox}[1][]{
colback=ForestGreen!15, 
colframe=ForestGreen!60,
    boxrule=0.5pt,
    arc=1.5mm,
    boxsep=1.2mm,
    left=2mm,right=2mm,top=2mm,bottom=2mm,
    #1
}

\newtcolorbox{answerbox}{
  enhanced,
  left=1.7mm,
  right=1.7mm,
  top=1.7mm,
  bottom=1.7mm,
  colback=gray!10,  
  colframe=gray!90, 
  boxrule=0pt,      
  leftrule=3pt,     
  sharp corners,    
  breakable         
}

\newcommand{\paratitle}[1]{\noindent\textbf{#1.}\quad}
\newcommand{\boxmargin}{1mm}
\newcommand{\tool}{\textsc{forge}\xspace}
\newcommand{\reports}{6,454\xspace}
\newcommand{\files}{81,390\xspace}
\newcommand{\vulnerabilities}{27,497\xspace}
\newcommand{\macrof}{86.1\%\xspace}
\newcommand{\precision}{95.6\%\xspace}
\newcommand{\recall}{78.4\%\xspace}
\newcommand{\CWE}{$CWE_s$\xspace}
\newcommand{\OWASP}{\textit{OWASP}\xspace}
\newcommand{\MITRE}{\textit{MITRE}\xspace}

\newcommand{\CVSS}{\textit{CVSS}\xspace}
\newcommand{\Kalpha}{Krippendorff's $\alpha$\xspace}
\newcommand{\kalpha}{$\text{k-}\alpha$\xspace}
\newcommand{\etal}{{\textit{et al.}}\xspace}
\newcommand{\eg}{{\textit{e.g.}}\xspace}
\newcommand{\ie}{{\textit{i.e.}}\xspace}

\begin{document}

\date{}

\title{\textsc{Forge}: An LLM-driven Framework for Large-Scale Smart Contract Vulnerability Dataset Construction}

\begin{abstract}

High-quality smart contract vulnerability datasets are critical for evaluating security tools and advancing smart contract security research.
Two major limitations of current manual dataset construction are
(1) labor-intensive and error-prone annotation processes limit the scale, quality, and evolution of the dataset, and (2) absence of standardized classification rules results in inconsistent vulnerability categories and labeling results across different datasets.
To address these limitations, we present \tool, the first automated approach for constructing smart contract vulnerability datasets.
\tool leverages an LLM-driven pipeline to extract high-quality vulnerabilities from real-world audit reports and classify them according to the Common Weakness Enumeration (CWE), the most widely recognized classification in software security.
\tool employs a divide-and-conquer strategy to extract structured and self-contained vulnerability information from unstructured audit reports. Additionally, it uses a tree-of-thoughts technique to classify the vulnerability information into the hierarchical CWE classification.
To evaluate \tool's effectiveness, we run \tool on \reports real-world audit reports and generate a dataset comprising \files solidity files and \vulnerabilities vulnerability findings across 296 CWE categories. Manual assessment of the dataset demonstrates high extraction precision and classification consistency with human experts (precision of \precision and inter-rater agreement \kalpha of 0.87). We further validate the practicality of our dataset by benchmarking 13 existing security tools on our dataset. The results reveal the significant limitations in current detection capabilities.
Furthermore, by analyzing the severity-frequency distribution patterns through a unified CWE perspective in our dataset, we highlight inconsistency between the current smart contract research focus and the priorities identified from real-world vulnerabilities. This analysis also reveals key differences from traditional software security concerns, providing practical implications for improving the smart contract security ecosystem.

\end{abstract}



\maketitle

\section{Introduction}
\label{sec:01_introduction}
Smart contracts on blockchain platforms have revolutionized digital asset management and decentralized applications~\cite{buterinNextgenerationSmartContract2014,zhengOverviewSmartContracts2020}, with the total value locked (TVL) exceeding \$117 billion~\cite{defillamaDefiLlama2025}. However, this rapid growth has been accompanied by significant security challenges.
In 2024, there were more than 760 on-chain security incidents, resulting in \$2.36 billion in losses~\cite{certikCertiKHack3dWeb32025}. These security incidents caused by evolving attack vectors~\cite{zhouSoKDecentralizedFinance2023} underscore the critical importance of robust vulnerability detection mechanisms in the smart contract ecosystem~\cite{chaliasosSmartContractDeFi2024a,wanSmartContractSecurity2021}.

High-quality vulnerability datasets are essential for developing security tools and advancing security research for smart contracts~\cite{zhengDAppSCANBuildingLargeScale2024,zhangDemystifyingExploitableBugs2023,sendnerLargeScaleStudyVulnerability2024}.
However, current approaches to smart contract vulnerability dataset construction face two critical limitations. \textbf{\textit{First, the manual process of dataset construction is labor-intensive and error-prone}}, which limits the scale, quality, and ability to keep up with the rapidly changing vulnerability landscape.
For instance, a recent smart contract vulnerability dataset construction~\cite{zhengDAppSCANBuildingLargeScale2024} required 44 person-months of manual effort to compile 1,618 vulnerabilities across 682 DApps, introducing potential quality concerns and making it difficult to evolve when new vulnerabilities emerge.
\textbf{\textit{Second, the absence of standardized classification rules}} leads to inconsistent vulnerability categories and labeling results across different datasets.
Existing datasets employ different vulnerability classifications, such as Smart Contract Weakness Classification (SWC)~\cite{swcSmartContractWeakness2024} or DASP10~\cite{nccgroupDASPTOP102021}, which may overlap or describe the same vulnerability but have different names~\cite{soudFlyOintmentEmpirical2023}.
For instance, a vulnerability labeled as ``Front-running''~\cite{nccgroupDASPTOP102021} might appear as ``Transaction Order Dependence (TOD)''~\cite{swcSWC114SmartContract2024} or ``Block Manipulation''~\cite{liStaticApplicationSecurity2024a} in another, despite describing the same underlying issue at different granularities.
The lack of unified vulnerability classification rules complicates the development and evaluation of vulnerability detection tools and leads to inconsistent vulnerability management practices~\cite{meunierFinalReport2nd1999}.

To address these key limitations, we present \tool, the first automated framework for constructing comprehensive and high-quality smart contract vulnerability datasets. \tool employs an LLM-driven pipeline to automatically extract vulnerability information from real-world audit reports~\cite{zhangWhenContractsMeets2024}, significantly reducing the manual effort required for dataset construction. To ensure standardized classification, we integrate the Common Weakness Enumeration (CWE)~\cite{mitreCWECWE2025} - the most widely recognized vulnerability classification in software security~\cite{liuNotEndStory2023,panFineGrainedCommitLevelVulnerability2023}. Specifically, to effectively extract vulnerability information from real-world audit reports, \tool implements a divide-and-conquer strategy through a map-reduce paradigm to effectively process lengthy audit reports and extract structured self-contained vulnerability information. To accurately classify the extracted vulnerabilities into the CWE classification, \tool introduces a tree-of-thoughts~\cite{yaoTreeThoughtsDeliberate2023a} reasoning technique that leverages in-context learning to classify vulnerabilities hierarchically to appropriate CWE categories.
Finally, \tool collects the source codes related to the extracted vulnerabilities and constructs CWE-labeled vulnerability entries within the vulnerability dataset.

To construct a comprehensive dataset and evaluate \tool's effectiveness, we first collect and filter out \reports smart contract audit reports from 47 security teams~\cite{etherscan.ioSmartContractsAudit2024}. Using the reports as input, \tool took only 229.5 hours to construct a large-scale dataset comprising \vulnerabilities vulnerability findings within \files Solidity files from real-world projects and covering 296 CWE categories. These files averaged 2,575 lines of code, with 59.0\% using the latest solidity compiler version (v0.8+). This efficient and automated process represents a significant improvement over existing manual dataset construction practices like DAppSCAN, which requires 44 person-months effort to build a dataset containing 39,904 files with 1,618 vulnerabilities from 25 categories~\cite{zhengDAppSCANBuildingLargeScale2024}. 
We evaluate the performance of \tool in extracting vulnerability-related information entities, achieving a Macro-F1 score of \macrof, with an average precision of \precision. Additionally, we assess the consistency of CWE classification between \tool and human experts, obtaining a high \Kalpha coefficient of 0.87~\cite{krippendorffContentAnalysisIntroduction2019}. Moreover, to validate the practicality of our dataset, we benchmark 13 widely used security tools. The results indicate their limited effectiveness, with the highest F1 score reaching 18.59\% and an average of 5.06\% across all tools.

To further provide insights with our dataset for security practitioners, we conduct an analysis of the smart contract vulnerability landscape, including visualizations of risk prioritization and comparative studies. By analyzing the severity-frequency distribution through a unified CWE classification, we reveal not only inconsistency between previous smart contract research focus and real-world sourced vulnerability priorities but also distinct characteristics compared to traditional software security concerns.

The key contributions of this research include:

\begin{itemize}[itemsep= 1 pt,leftmargin=*,topsep = 1pt, parsep = 1 pt]
    \item We introduce \tool, the first LLM-based framework designed to automatically construct smart contract vulnerability datasets by extracting vulnerability information from audit reports and classifying it into the CWE classification.
    \item We used \tool to construct a dataset that includes \vulnerabilities CWE-annotated vulnerability findings from \files real-world Solidity files. This dataset features an average of 2,575 lines of code per project, with 59.0\% of projects using the latest Solidity compiler versions (v0.8+). 
    \item We highlight the limitations of existing security tools and conduct an empirical analysis that combines the frequency and severity of each vulnerability from a CWE perspective to derive insights from previous research on both smart contracts and traditional software.
    \item We have made our dataset, experimental results, and the source code of \tool available at \href{https://github.com/shenyimings/FORGE-Artifacts}{\color{NavyBlue}{\url{https://github.com/shenyimings/FORGE-Artifacts}}}.
\end{itemize}

\section{Background and Motivation}
\label{sec:02_background}

\begin{figure*}[!th]
    \centering
    \includegraphics[width=1.0\textwidth, clip]{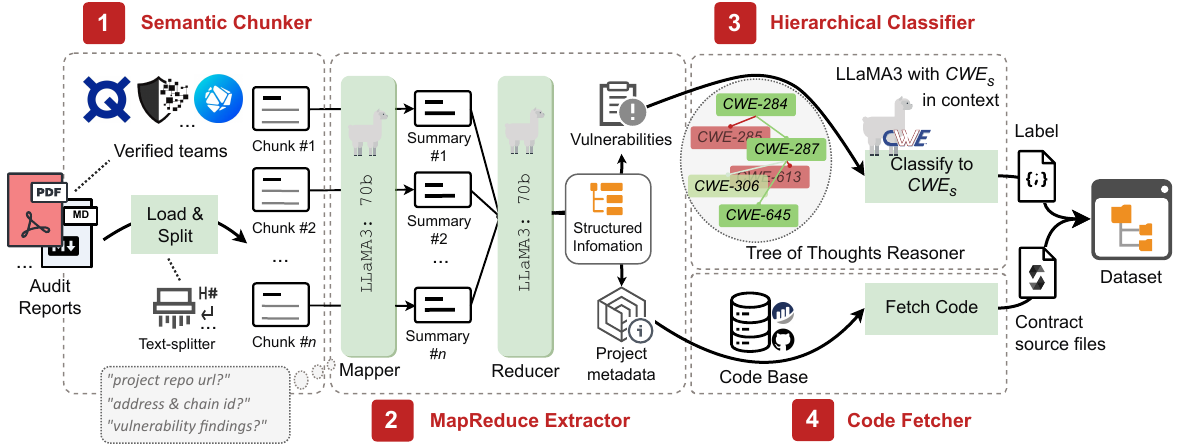}
    \Description{Overview of \tool framework}
    \vspace{-1.8em}
    \caption{Overview of \tool framework.}
    \label{fig:architecture}
    \vspace{-0.8em}
    
\end{figure*}

\subsection{Common Weakness Enumeration (CWE)}

Common Weakness Enumeration (CWE) is a dictionary of software and hardware vulnerabilities, regularly maintained by the community to reflect emerging security issues~\cite{mitreCWECWE2025,soudFlyOintmentEmpirical2023}. 

\textit{CWE-1000: Research Concepts}~\cite{mitreCWECWE1000Research2025} view provides a comprehensive classification system that follows a deep tree structure with four abstraction levels that organize weaknesses from abstract concepts to specific implementations: \textit{Pillar}, \textit{Class}, \textit{Base}, and \textit{Variant}. 
At the highest level, there are ten \textit{Pillar} level vulnerability categories, representing fundamental vulnerabilities that cannot be further abstracted. For example, \textit{CWE-284: Improper Access Control}~\cite{mitreCWECWE284Improper2025} is one of the ten \textit{Pillar} level weaknesses, with \textit{CWE-287: Improper Authentication}~\cite{mitreCWECWE287Improper2025} as its \textit{Class}-level child. Moving down the hierarchy, the \textit{Base} level introduces specific behaviors and properties as illustrated by \textit{CWE-295: Improper Certificate Validation}~\cite{mitreCWECWE295Improper2025}. This specificity is further refined at the \textit{Variant} level, where technology-specific details and implementations are addressed, as demonstrated by \textit{CWE-298: Improper Validation of Certificate Expiration}~\cite{mitreCWECWE298Improper2025}.

The \textit{CWE Classification} process classifies diverse real-world security issues to standardized CWE entries. This process identifies and codifies the underlying flaws or errors responsible for exploitable security risks~\cite{mitreCWECVECWE2025}. The hierarchical structure of CWE allows for the classification of nearly any software vulnerability to an appropriate entry. Prior studies~\cite{soudFlyOintmentEmpirical2023, ruggieroSoKUnifiedData2024, zhouStateEthereumSmart2022a} and audit reports \eg~\cite{chainsulting1inchV2Audit2020} have adopted CWE as an identifier for smart contract vulnerabilities.

\subsection{Smart Contract Audit Reports}
\label{sec:bg-artifacts}

Smart contract auditing is a rigorous security assessment process conducted by professional audit teams (\eg, Etherscan-verified audit teams~\cite{etherscan.ioSmartContractsAudit2024}) to identify security issues and document vulnerabilities in detailed audit reports~\cite{hederaWhatSmartContract2025}. 
These expert-validated reports not only detail vulnerability attack vectors but also provide essential context about their discovery and potential impact, making them ideal sources for systematic vulnerability analysis~\cite{zhangDemystifyingDetectingCryptographic2024,huangRevealingHiddenThreats2024a}. 
They aim to identify CWE-related vulnerabilities in smart contracts and may recommend how to fix them.

\subsection{LLMs for Dataset Construction}
\label{sec:challenges}
Large Language Models (LLMs) are advanced machine learning models trained on extensive datasets of text and code, capable of understanding and generating human-like language across diverse domains~\cite{NEURIPS2022_8bb0d291}. Notable examples include \textit{GPT-4}~\cite{openaiGPT42023}, \textit{Claude3.5}~\cite{anthropicIntroducingNextGeneration2024}, and \textit{Llama3}~\cite{aiIntroducingMetaLlama2024}. 
These powerful models have demonstrated remarkable capabilities in natural language processing~\cite{tanLargeLanguageModels2024,wangChatGPTGoodNLG2023}, program understanding~\cite{zhengUnderstandingLargeLanguage2024}, and security analysis~\cite{sunLLM4VulnUnifiedEvaluation2024,liuNotEndStory2023}.

Existing manual dataset construction processes are highly labor-intensive and error-prone, which largely limits the scale, quality, and evolution of the dataset.
The promising capabilities of LLMs in program comprehension and information extraction~\cite{liuExploringChatGPTsCapabilities2023} demonstrate the potential to reduce the manual effort required by the current dataset construction processes.
This motivates us to investigate whether an LLM-driven pipeline can be leveraged to achieve automated vulnerability dataset construction.

Applying LLMs to build CWE-labeled smart contract vulnerability datasets from complex real-world audit reports introduces three key challenges:
\textit{(1) Complexity of audit reports.} For instance, the Trail of Bits' audit report for Uniswap v4 Core~\cite{remieUniswapV4Core2024} spans 63 pages, documenting project metadata and vulnerability findings alongside extensive audit disclaimers, testing methodology descriptions, and code architecture explanations. Such complexity necessitates a systematic approach to extract and aggregate vulnerability information efficiently.
\textit{(2) Complexity of the CWE hierarchy.} With over 900 weakness types across multiple abstraction levels and dimensions~\cite{mitreCWECWE1000Research2025}. Traditional approaches such as keyword match~\cite{mitre2024CWETop2024} and knowledge graph~\cite{hanDeepWeakReasoningCommon2018} often lead to inconsistent or superficial results\cite{liuNotEndStory2023,simsekPosterAnalyzingCorrecting2024}. 
\textit{(3) Limitations of LLMs.} LLMs exhibit constraints in handling domain-specific technical tasks. Their limited context window and the ``lost in the middle'' phenomenon~\cite{liuLostMiddleHow2024} impair the processing of lengthy audit reports and lead to overlooked content details~\cite{jiangStructGPTGeneralFramework2023a}. Moreover, their inherent tendency to generate hallucinated content~\cite{zhengUnderstandingLargeLanguage2024} could compromise the precision of vulnerability classification.

\section{\tool Framework}

To automate smart contract vulnerability dataset construction and to improve scaling, quality, and evolution, we present \tool, an end-to-end framework that leverages LLMs to build CWE-labeled datasets from real-world audit reports.

\subsection{Overview}

The workflow of \tool is outlined in Figure~\ref{fig:architecture}. The process begins with the \textit{Semantic Chunker}, which takes an audit report as input and segments it into self-contained chunks based on semantic boundaries in the report file. The \textit{MapReduce Extractor} then processes individual chunks to extract vulnerability information (map phase) and aggregate the results into structured summaries (reduce phase). The \textit{Hierarchical Classifier} employs the tree-of-thoughts~\cite{yaoTreeThoughtsDeliberate2023a} approach to systematically classify vulnerabilities into the \CWE (software-related entries of CWE) hierarchy, utilizing in-context learning~\cite{dong-etal-2024-survey-icl} with domain knowledge injection at each decision point. Finally, the \textit{Code Fetcher} module retrieves and integrates relevant smart contract source files, complementing the extracted vulnerability information to produce structured and classified vulnerability findings for vulnerability dataset construction.

\subsection{Phase 1: Semantic Chunker}
\label{sec:chunker}

To address the token length limitations of LLMs, while preserving semantic integrity, we first apply a \textit{Semantic Chunker} module. This splits long-form audit reports into self-contained, semantically coherent chunks by segmentation and chunk size optimization.
Specifically, a chunk is a text segment that (1) preserves complete semantic information related to a vulnerability and (2) length falls within the LLM's processing threshold.

To split the audit report into self-contained chunks, \tool first converts the input report file into a unified markdown format using \texttt{PyMuPDF}~\cite{pymupdfPyMuPDF2025}, preserving the file structure of headings and content. Then, it operates in the following steps:
\textit{First}, the document's inherent structure is leveraged to segment content along natural semantic boundaries. These boundaries include document headings, paragraph breaks, and Unicode-defined text units (from grapheme clusters to sentences) to preserve the semantic hierarchy of the source document.
\textit{Then}, a size verification ensures all chunks remain within the LLM's token limit. If necessary, oversized chunks are further divided at appropriate boundaries recursively.

\subsection{Phase 2: MapReduce Extractor}

To systematically process diverse and lengthy audit reports, \tool employs the MapReduce Extractor following the semantic chunking phase. This module adopts a divide-and-conquer strategy, which first extracts structured project metadata and vulnerability information from these chunks (\ie, the map stage) and then merges the vulnerability information among chunks (\ie, the reduce stage).
Specifically, the project metadata encompasses blockchain network, on-chain address, GitHub URL, and commit ID, which is necessary for accessing the source code associated with identified vulnerabilities. The vulnerability information includes the title, description, severity, and location, elucidating the vulnerability's attack vector, prerequisites, and potential impacts of exploitation.
In the map stage, \tool extracts information for each chunk $c_i$:

\vspace{-0.5em}
\begin{equation}
s_i = f_{\text{map}}(c_i, \mathcal{Q}; \theta)
\end{equation}

\noindent where $\mathcal{Q}$ denotes the information types that the LLM needs to extract, including project metadata and vulnerability findings. The LLM-powered mapping function $f_{\text{map}}$, parameterized by $\theta$, performs extraction of input chunks $c_i$, producing outputs $\{s_1, \ldots, s_N\}$ conforming by $\mathcal{Q}$.

In the reduce stage, the final structured information is generated through:

\vspace{-0.5em}
\begin{equation}
a = f_{\text{reduce}}(\{s_1, \cdots, s_K\}, \mathcal{P}, \mathcal{V}; \theta)
\end{equation}

\noindent where $\mathcal{P}$ denotes project metadata comprising source code identifiers (including GitHub repository URLs with specific commit hashes or on-chain contract addresses), and $\mathcal{V}$ represents vulnerability attributes containing title, description, severity level, and affected code locations. 
The $f_{\text{reduce}}$ operation first concatenates map results $\{s_1, \cdots, s_K\}$ until reaching the \texttt{chunk\_length} threshold defined in Section~\ref{sec:chunker}, then instruct LLM parameterized by $\theta$ to deduplicate and merge the map results to a structured JSON output $a$. Figure~\ref{fig:structured} illustrates this structured output schema.

\begin{figure}[htb]
\setlength{\abovecaptionskip}{0.2cm}
\begin{lstlisting}[language=JSON,firstnumber=1]
project_info: 
    url: string,
    commit_id: string,
    address: string,
    chain: string
findings: 
    id: int,
    title: string,
    description: string, 
    severity: enum(critical, high, medium, low, info),
    location: string
\end{lstlisting}
\caption{Structured information}
\vspace{-0.5em}
\label{fig:structured}
\end{figure}

In cases where the combined length of map summaries ($\Sigma s_i$) exceeds the model's context window limit \texttt{chunk\_length}, the MapReduce extractor employs a two-stage reduction strategy. First, it partitions the $N$ summaries into $K$ groups, where each group's total length remains within \texttt{chunk\_length}. The system then performs $K$ independent reduce operations, each producing a partially structured result containing project metadata and vulnerability information.

\begin{figure}[htbp]
    \centering
    \includegraphics[width=0.85\linewidth]{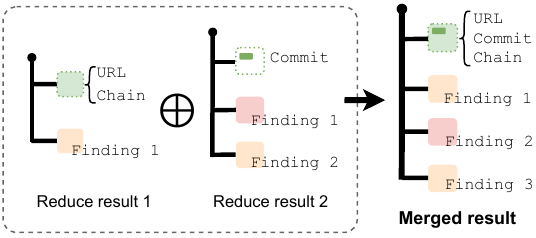}
    \vspace{-0.5em}
    \caption{A diagram of JSON merge operation}
    \vspace{-0.8em}
    \label{fig:merge-example}
    \Description{A diagram of JSON merge operation}
\end{figure}

To merge these $K$ partial results into a final output, we define an operator $\oplus$ that combines both project metadata ($\mathcal{P}$) and vulnerability information ($\mathcal{V}$) from two reduced JSON results in $a_i, a_j$: 

\begin{equation}
a_i \oplus a_j = \{\mathcal{P}_i \cup \mathcal{P}_j, \mathcal{V}_i \cup \mathcal{V}_j\}
\end{equation}

As illustrated in Figure~\ref{fig:merge-example}, the merge operation discards empty fields and combines non-conflicting fields. 
In cases of conflicts, the system adopts information from chunks that appear \textit{earlier} in the original document sequence. This choice is based on experimental observations indicating that project metadata, \eg URLs, commit IDs, and addresses, typically appears in the earlier sections of audit reports. Consequently, the information present in the earlier chunks is considered more reliable. It has a lower likelihood of hallucination by the LLM compared to the information found in later chunks.

Our map-reduce-based design ensures the completeness and consistency of extracted information while handling long-form reports that exceed the model's context limitations.

\subsection{Phase 3: Hierarchical Classifier}

After completing smart contract vulnerability information extraction, we then classify the structured vulnerability findings into CWE categories.
Given the extensive parent-child structure of CWE classifications as detailed in Section~\ref{sec:challenges}, our \textit{Hierarchical Classifier} module integrates an LLM-driven tree-of-thoughts (ToT)~\cite{yaoTreeThoughtsDeliberate2023a} reasoning process to navigate the complex hierarchical structure of CWE, and utilizes In-Context Learning (ICL)~\cite{dong-etal-2024-survey-icl} for CWE knowledge injection for accurate classification.

\subsubsection{Pre-processing}
To better characterize software-level vulnerabilities relevant to smart contracts, we refine the \textit{CWE-1000: Research Concepts} view~\cite{mitreCWECWE1000Research2025}, which contains both hardware and software vulnerability categories. Specifically, we identified and filtered out 108 hardware-related entries (\eg, \textit{CWE-1192: Improper Identifier for IP Block used in System-On-Chip}~\cite{mitreCWECWE1192Improper2025}) that are irrelevant to software security assessment. We denote this refined subset as $CWE_{s}$, which forms the foundation for the \textit{Hierarchical Classifier}. Unless otherwise specified, all subsequent references to CWE categories in this paper refer to entries within this software-focused subset.

Additionally, following guidance from the CWE community~\cite{mitreCWECVECWE2025}, we labeled each entry in $CWE_{s}$ with mapping notes indicating whether the entry is suitable as a final category for smart contract vulnerability. These labels guide our classifier by signaling which \CWE nodes can serve as valid endpoints during the classification process, enabling proper fallback to higher-level categories when more specific classifications are not appropriate.
The detail of this subset with mapping notes is provided in our online repository.

\subsubsection{Tree-of-Thoughts (ToT) Reasoning}
Our classification approach leverages the inherent tree structure of CWE, where vulnerabilities are organized in a hierarchical taxonomy from pillar-level categories to concrete leaf nodes~\cite{mitreCWECWE1000Research2025}. 
By adopting a tree-of-thoughts (ToT) reasoning framework~\cite{yaoTreeThoughtsDeliberate2023a}, \tool can systematically explore classification decisions from general to specific CWE categories while maintaining previous decision paths.

Our method transforms the vulnerability classification challenge into a guided tree node search problem, where the LLM -- with \CWE knowledge in context -- serves as a pathfinder that traverses through the \CWE hierarchy.
At each level, LLM considers the vulnerability information alongside the local structure of candidate \CWE categories, determining whether to advance to more specific categories or fall back to previous nodes when appropriate.

\begin{algorithm}
\caption{Tree-of-Thoughts Reasoner}
\label{alg:cwe-tot}
\begin{algorithmic}[1]
\Require vuln\_info; llm; $CWE_s$; $k$, $l$
\State \textbf{global} path $\gets []$
\Function{Classify}{node, l}
    \State \textit{children} $\gets$ \Call{GetChildren}{$CWE_s$, node}
    \If{children is empty} \Return \EndIf
    \State \textit{fallback\_node} $\gets []$
    \If{node is MappingAllowd} fallback\_node $\gets$ node \EndIf
    \State \textit{prompt} $\gets$ \Call{ConstructPrompt}{vuln\_info, fallback\_node, children, k}
    \State selected\_node(s) $\gets$ \Call{Parse}{llm(\textit{prompt})}
    \If{fallback\_node in selected\_node(s)} \Return \EndIf
    \State path.append([(l, selected\_node(s))])
    \State \Call{Classify}{selected\_node(s), l+1}
\EndFunction
\State \textbf{initialize} vuln\_info, llm, $CWE_s$, k
\State \Call{Classify}{root of $CWE_s$, 0}
\State \Return path
\end{algorithmic}
\end{algorithm}

We formalize the classification process as a sequence of structured reasoning steps, as presented in Algorithm~\ref{alg:cwe-tot}. It takes the vulnerability information $vuln$, the language model $llm$, the pruned CWE hierarchy $CWE_s$, the number of most relevant child nodes to be selected at each level $k$ and the current level in the hierarchy $l$ as inputs. The output is the classification $path$.

The algorithm performs a recursive traversal of the $CWE_s$ hierarchy, starting from the root node and progressively classifying the vulnerability into more specific categories.
At each level $l$, $llm$ functions as a classifier, taking prompt as input: (1) the vulnerability title and description, (2) the current set of candidate child nodes with their descriptions, and (3) a parameter $k$ specifying the number of relevant nodes to select. $llm$ leverages its semantic understanding capabilities to reason between the vulnerability information and each candidate node's characteristics, outputting the $k$ most relevant child \CWE nodes.

\subsubsection{Fallback Strategy}
In some cases, the most appropriate $CWE_s$ category for a vulnerability may not be a leaf node but a more abstract node closer to the root. To handle this, we introduce a fallback strategy. For $CWE_s$ nodes that are labeled as mapping allowed, we add them to the $fallback\_node$ list in addition to their child nodes (lines 8-10 in Algorithm~\ref{alg:cwe-tot}). If the $ llm$ selects a fallback node, the recursion terminates, and the final classification path is returned (lines 13-15). Otherwise, the recursion continues to level $l+1$ if child nodes are selected by $llm$ (lines 16-17).

\begin{figure}[htbp]
    \centering
    \vspace{-1em}
    \includegraphics[width=0.9\linewidth]{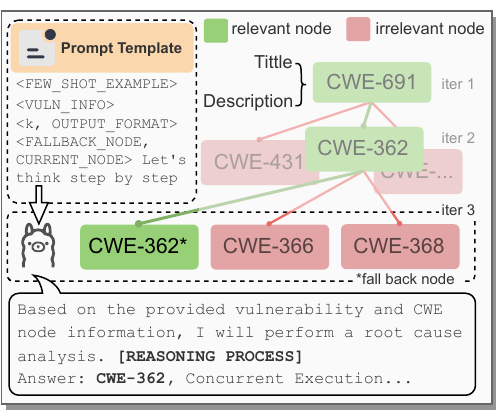}
    \vspace{-0.5em}
    \caption{An example of ToT workflow}
    \vspace{-0.2em}
    \label{fig:tot-example}
    \Description{An example of ToT classification workflow}
    
\end{figure}

Figure~\ref{fig:tot-example} shows an example of the \CWE classification workflow. For a vulnerability finding related to a marketplace contract affected by potential front-running manipulations due to the absence of a minimum swap output value enforcement mechanism, the classification proceeds as follows.
With $k=1$, in the first iteration, it is classified into \textit{CWE-691: Insufficient Control Flow Management}~\cite{mitreCWECWE691Insufficient2025} out of the 10 pillar categories. In the second iteration, from the child nodes of \textit{CWE-691}, the vulnerability is further classified into \textit{CWE-362: Concurrent Execution using Shared Resource with Improper Synchronization (`Race Condition')}~\cite{mitreCWECWE362Concurrent2025}. \textit{CWE-362} is labeled as mapping allowed, indicating that it is a suitable category for smart contract vulnerability mapping. Therefore, in the third iteration, \textit{CWE-362} is added to the selection list as a fallback node along with its child nodes. LLM selects CWE-362 again, and the iteration ends. The final classification path is \textit{CWE-691}$\to$\textit{CWE-362}.

\subsection{Phase 4: Code Fetcher}

The metadata of target projects obtained from the extractor is then processed by the \textit{Code Fetcher}. This retrieves the corresponding smart contract source code from trusted public repositories -- GitHub and blockchain explorers such as Etherscan~\cite{etherscan.ioEthereumETHBlockchain2025} and Bscscan~\cite{bscscanBNBSmartChain2025}).
The process of dealing with on-chain code is straightforward due to blockchain's immutable nature.  Source code at a specific contract address directly corresponds to the version affected by the reported vulnerability. For smart contract projects hosted on Github repositories, developers typically create an \textit{audit} branch for review~\cite{uniswapUniswapV4coreAudit2024}, which undergoes a thorough examination by security auditors. Subsequently, developers implement fixes through new commits. To maintain precise vulnerability information tracking, our \textit{MapReduce} extractor module captures the \texttt{commit-id} (typically in the form of a hash or a descriptive name) of the audited codebase, allowing the \textit{Code Fetcher} module to retrieve the exact version of the source code files.
Finally, the \textit{CodeFetcher}  combines the collected source code files and the extracted vulnerability information into vulnerability entities within the constructed dataset.

\section{Evaluation}
In this section, we intend to answer the following key research questions (RQs):

\begin{itemize}[leftmargin=0.7cm]
    \item \textbf{RQ1. Can \tool effectively generate a large-scale smart contract vulnerability dataset from real-world audit reports? }\xspace
    We run \tool on \reports audit reports to construct a dataset, analyzing the statistics and comparing them with existing datasets.
    \item \textbf{RQ2. How does \tool perform in vulnerability information extraction? }\xspace
    We assess the effectiveness of \tool framework in information extraction.
    \item \textbf{RQ3. How consistently does \tool perform in vulnerability classification against human experts? }\xspace
    We assess the effectiveness of \tool framework for vulnerability classification.
    \item \textbf{RQ4. How practical is the \tool dataset for evaluating current smart contract vulnerability detection tools? }\xspace
    We leverage our large-scale dataset to assess the efficacy and limitations of existing security tools, highlighting the practicability of the dataset.
\end{itemize}

\paratitle{Data Collection}
To curate our experimental dataset, we collect 10,312 publicly available audit reports from 47 renowned auditing teams verified through Etherscan~\cite{etherscan.ioSmartContractsAudit2024}.
Since our dataset construction requires both vulnerability information and the corresponding source code, we filter out audit reports where the source code is not publicly accessible. We apply a regular expression matching script to identify valid on-chain addresses and source code repository URLs, ultimately obtaining \reports documents for further analysis. 

\paratitle{Experiment Setup}
We conducted our experiments on a CentOS 7.9 server equipped with 128 Intel(R) Xeon(R) Platinum 8376H CPUs @ 2.60GHz, 512GB RAM, and 2 NVIDIA A800 80GB PCI GPUs. We employ \textit{Llama3:70b-instruct-q8\_0} as the foundation model for the \textit{MapReduce Extractor} and \textit{Hierarchical Classifier} modules, with a chunk length of 4,096 tokens, the number of a most relevant child node $k$ set to 1 and a default temperature setting of 0.8~\cite{andreiLlamacpp}.

\subsection{RQ1: Effectiveness of \tool}
\label{sec:rq1}

We begin by running \tool on our previously collected \reports audit reports. This took a total of 229.5 hours, with an average processing time of 127.8 seconds per report. This includes 45.0 seconds for completing a vulnerability information extraction task and 18.3 seconds for each vulnerability classification. 

\begin{table}[!ht]
    \vspace{-0.8em}
    \centering
    \caption{Overview of our \tool dataset}
    \vspace{-0.8em}
    \begin{tabular}{lc}
    \toprule
        \textbf{Statistics} & \textbf{Numbers} \\
    \midrule
        Total audit reports & \reports \\
        Total DApp projects & 6,579 \\
        Total solidity files & \files \\
        Average solidity files in a project & 12 \\
        Average line of code in a project & 2,575 \\
    \midrule
        Compiler Version 0.4+ & 270 \\
        Compiler Version 0.5+ & 478 \\
        Compiler Version 0.6+ & 1,524 \\
        Compiler Version 0.7+ & 360 \\
        Compiler Version 0.8+ & 3,791 \\
        Other Compiler Version & 31 \\ 
    \midrule
        Total vulnerability findings & \vulnerabilities \\
    \bottomrule
    \end{tabular}
    \vspace{-0.8em}
    \label{tab:dataset_overview}
\end{table}

The dataset built by \tool encompasses \files Solidity files, covering \vulnerabilities vulnerabilities with CWE labels. Essential information for each project is stored in \reports JSON files. Table \ref{tab:dataset_overview} outlines the key parameters of our dataset, highlighting that each project contains an average of n Solidity smart contract files, with a mean of 2,575 lines of code per project. Notably, the most prevalent compiler version in our dataset is 0.8+ (59.0\%).
Compared to the widely-used SmartBugs dataset~\cite{durieuxEmpiricalReviewAutomated2020}, where contracts average only 204 lines of code and over 90\% use outdated compiler versions (0.4+)~\cite{zhengDAppSCANBuildingLargeScale2024}, \textbf{our real-world sourced dataset exhibits significantly higher complexity and comprehensiveness}. Similarly, it also shows \textbf{a substantial increase in scale and coverage} compared to human-annotated datasets \eg DAppSCAN, which contains 39,904 files with 1,618 vulnerabilities, with 35.2\% of them using compiler 0.8+~\cite{zhengDAppSCANBuildingLargeScale2024}. 

\begin{figure}[t]
    \centering
    \includegraphics[width=1\linewidth]{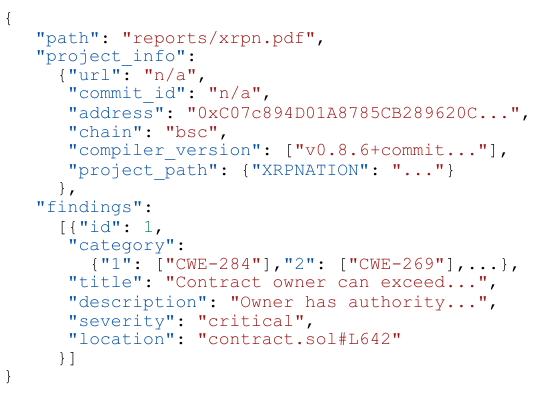}
    \vspace{-2em}
    \caption{An example record of \tool dataset}
    \label{fig:dataset_example}
    \vspace{-1em}
\end{figure}
Figure \ref{fig:dataset_example} showcases an example of a typical record in our dataset. Each entry is comprised of several key fields that provide comprehensive information about the audited smart contract and its vulnerabilities. The "\texttt{path}" field identifies the corresponding reports, while the "\texttt{project\_info}" field captures essential metadata about the DApp, including its URL, commit ID, contract address, blockchain network, compiler version, and Solidity file paths. The "\texttt{findings}" field is an array where each element represents a distinct vulnerability. These vulnerability entries are further detailed with a unique identifier ("\texttt{id}"), a CWE hierarchical classification ("\texttt{category}"), a concise description ("\texttt{title}"), an in-depth explanation ("\texttt{description}"), an assessed impact level ("\texttt{severity}"), and the specific code location of the vulnerability ("\texttt{location}")

\begin{answerbox}
\textbf{Answer to RQ1:} \tool approach has built the most comprehensive smart contract vulnerability dataset compared to previous attempts, such as Smartbugs and DAppSCAN, in both scope and completeness.
\end{answerbox}

\subsection{RQ2: Performance of Information Extraction}
\label{sec:rq2}

To evaluate the performance of \tool in information extraction, we first adopt a random sampling method based on confidence intervals from the dataset built in RQ1, following the practices of prior works~\cite{liuCharacterizingTransactionrevertingStatements2022,yangDefinitionDetectionDefects2023a}. We set the confidence level to 95\% and the confidence interval to 10. Using the calculation method implemented in~\cite{calculator.netSampleSizeCalculator2025}, we randomly sampled 96 samples from the dataset. Two authors of this paper independently match the results produced by \tool with original audit reports. Any discrepancies in this labeling were resolved through discussion with a third author.

Following the evaluation method in~\cite{dagdelenStructuredInformationExtraction2024}, we then calculate precision (P) and recall (R) for each entity type. Precision measures the accuracy of the extracted information, while recall measures the comprehensiveness of the extraction. These metrics are calculated by $\frac{\text{\# correct entities extracted}}{\text{\# entities extracted}}$ and $\frac{\text{\# correct entities extracted}}{\text{\# entities in ground truth}} $, respectively. The F1-score computes using the formula: $\frac{2\times(P\times R)}{P + R}$

\begin{table}[ht]
\centering
\vspace{-0.5em}
\caption{Information extraction performance of \tool.}
\vspace{-0.8em}
\setlength{\tabcolsep}{4pt}
\begin{tabular}{lccc}
\toprule
\textbf{Entity} & \textbf{Precision(\%)} & \textbf{Recall(\%)} & \textbf{F1(\%)} \\
\midrule
on-chain address & 92.6 & 73.5 & 82.0 \\
chain & 100.0 & 82.3 & 90.3 \\
URL & 100.0 & 76.7 & 86.8 \\
commit ID & 100.0 & 78.9 & 88.2 \\
\midrule
vulnerability finding & 91.7 & 73.2 & 81.4 \\
severity & 88.3 & 81.9 & 85.0 \\
location & 96.6 & 82.5 & 89.0 \\
\midrule
\textbf{Average} & 95.6 & 78.4 & \textbf{86.1} \\
\bottomrule
\end{tabular}
\vspace{-0.5em}
\label{tab:extraction_performance}
\end{table}

Table~\ref{tab:extraction_performance} presents the information extraction performance of \tool, structured with entity types divided into project metadata (on-chain address, chain, URL, commit ID) and vulnerability-related (vulnerability finding, severity, location), showing precision, recall, and F1-score for each category. We then compute \tool's overall metrics as a weighted average across all entity types. \textbf{We find that \tool achieves an overall precision of \precision, with a recall of \recall. The average F1-score (Macro-F1) is \macrof.}

Our analysis of incorrect samples reveals several scenarios for improvement. Some reports present complex information (\eg vulnerability codes) as images. In contrast, others employ checklist formats with icons or color highlights to mark inspection results, which poses challenges for single-modal models in detecting vulnerability information. These modality constraints impact extraction comprehensiveness but do not impact the reliability of successfully identified vulnerabilities.

\begin{answerbox}
\textbf{Answer to RQ2:} Our \tool framework achieved an overall F1 rate of \macrof with a precision of \precision, highlighting its performance in extracting project and vulnerability information.
\end{answerbox}

\subsection{RQ3: \tool's Classification v.s. Human Experts}
\label{sec:rq3}

To answer RQ3, we evaluate the consistency of \tool's vulnerability classification against human expert judgments. 
Determining the category of a vulnerability with limited information is a subjective and experience-based task, which may lead to varying results even for professional security auditors~\cite{mitreCWECVECWE2025}. 
For instance, the common smart contract vulnerability \textit{Authorization through `tx.origin'} has been categorized differently across security standards: SWC~\cite{swcSWC115SmartContract2020} categorized it under \textit{CWE-477: Use of Obsolete Function}~\cite{mitreCWECWE477Use2025} due to its outdated and deprecated nature, while the EthTrust specification~\cite{nevileEEAEthTrustSecurity2023} classified it under \textit{CWE-284: Improper Access Control}~\cite{mitreCWECWE284Improper2025} because it can facilitate unauthorized access. Such discrepancies are particularly pronounced when analyzing vulnerabilities with limited contextual information.
Thus, we employ \Kalpha (\kalpha)~\cite{ahmedCanLLMsReplace2024, krippendorffContentAnalysisIntroduction2019}, a robust reliability coefficient widely used in content analysis with multiple evaluators, to quantitatively assess the consistency between \tool's classifications with human experts, instead of using human labeling results as ground truth.

Following the evaluation methodology from RQ2, we set a confidence level of 95\% and a confidence interval of 10, randomly selecting 96 samples from the \vulnerabilities vulnerabilities annotated with CWE categories. Two authors of this paper independently analyzed the description and code of each sample vulnerability, mapping it to the closest category in the CWE classification system based on its root cause. These ground truths were then compared with the ultimate classification results of the \tool framework. 

We calculated \kalpha using Equation~\ref{eq:k-alpha}, where $D_o$ represents the observed disagreement, $D_e$ denotes the expected disagreement by chance, $n_{ijk}$ is the number of disagreements between coders i and j on coding unit k, $d_{ijk}$ is the observed distance between coders i and j on coding unit k, and $E(d_{ijk})$ is the expected random disagreement distance.
\vspace{-0.1cm}
\begin{equation}
    \alpha = 1 - \frac{D_o}{D_e} = 1 - \frac{\sum_{i<j} \sum_{k} n_{ijk} d_{ijk}}{\sum_{i<j} \sum_{k} n_{ijk} E(d_{ijk})} \label{eq:k-alpha}
\end{equation}

The resulting \kalpha coefficient is \textbf{0.87}, above the threshold of 0.80 suggested by Krippendorff~\cite{marziKAlphaCalculatorKrippendorffs2024} for drawing reliable conclusions. \textbf{This high level of agreement indicates that our LLM-based classification approach achieves a substantial level of reliability.}

There may be some inconsistencies in \tool. We have open-sourced our dataset on GitHub to enable community engagement, allowing developers and researchers to report issues and contribute to its ongoing improvement. We analyzed the inconsistent results we found and discovered three main reasons causing these inconsistencies.

\begin{itemize}[leftmargin=*]
    \item \textbf{Some vulnerability descriptions are overly general}, requiring experience-based inference to accurately assign the corresponding CWE type, leading to inconsistency between human and LLM judgments.
    \item In \textbf{complex cases involving multiple vulnerability categories}, human experts tend to select more specific subcategories, while LLM tends to choose more generic parent categories.
    \item The \textbf{ambiguity of certain low-level CWE categories} introduces challenges for consistent classification. For example, distinguishing between \textit{CWE-755: Improper Handling of Exceptional Conditions}~\cite{mitreCWECWE755Improper2025} and \textit{CWE-754: Improper Check for Unusual or Exceptional Conditions}~\cite{mitreCWECWE755Improper2025} requires careful consideration due to their conceptual similarity.

\end{itemize}

\begin{answerbox}
\textbf{Answer to RQ3:} \tool achieved a 0.87 \kalpha coefficient, indicating substantial agreement between \tool and human experts in CWE classification of smart contract vulnerabilities. 
\end{answerbox}

\subsection{RQ4: Practicality of \tool dataset}

We assess the practicality of the \tool dataset by evaluating how existing smart contract vulnerability detection tools perform when applied to our CWE-classified vulnerabilities derived from real-world audit reports. 
To conduct this assessment, we employ SmartBugs, a framework that integrates various tools to analyze smart contracts~\cite{durieuxEmpiricalReviewAutomated2020}. The selection criteria for the tools included the requirement that tools support source code as input, be automated, and be capable of detecting at least one type of CWE vulnerability. Based on these criteria, we selected 13 representative vulnerability detection tools listed in Table~\ref{tab:tool_comparison} with diverse detection techniques.

Two authors independently map the vulnerabilities each tool can detect to CWE categories, following the vulnerability mapping guidelines provided in SmartBugs-Wiki~\cite{smartbugsVulnerabilitiesMapping2020}. Any discrepancies in this mapping were resolved through discussion with a third author. Based on this~\href{https://github.com/shenyimings/FORGE-Artifacts/blob/main/evaluation/RQ4/tool_classifications.csv}{mapping}, we run the selected tools on the \tool dataset with a default 300-second timeout and collect their detection results. We standardize evaluation at the contract level: if a tool detects any vulnerability within the contract that contains labeled vulnerabilities, we count it as a true positive (TP). We calculate precision, recall, and F1-score for each tool following the same methodology as~\cite{zhengDAppSCANBuildingLargeScale2024, chenNumScoutUnveilingNumerical2025}, as shown in Table \ref{tab:tool_comparison}.

\begin{table}[htbp]
\centering
\caption{Analysis results of existing detection tools}
\vspace{-0.8em}
\resizebox{1.0\columnwidth}{!}{
\begin{tabular}{lcccccc}
\toprule
\textbf{Tool} & \textbf{TP} & \textbf{FP} & \textbf{FN} & \textbf{P(\%)} & \textbf{R(\%)} & \textbf{F(\%)} \\
\midrule
Confuzzius~\cite{torresConFuzziusDataDependencyAware2021} & 13 & 462 & 2,250 & 2.74 & 0.57  & 0.95  \\
Conkas~\cite{conkasNvelosoConkasEthereum2021} & 2 & 51 & 677 & 3.77 & 0.29 & 0.55 \\
Honeybadger~\cite{torresArtScamDemystifying2019} & 0 & 7 & 52 & 0.00 & 0.00 & 0.00  \\
Maian~\cite{nikolicFindingGreedyProdigal2018} & 0 & 10 & 1,124  & 0.00 & 0.00  & 0.00 \\
Manticore~\cite{mossbergManticoreUserFriendlySymbolic2019} & 0 & 0 & 992 & 0.00 & 0.00 & 0.00 \\
Mythril~\cite{consensysConsensysMythrilMythril2018} & 0 & 33 & 4,383 & 0.00 & 0.00 & 0.00 \\
Osiris~\cite{torresOsirisHuntingInteger2018b} & 1 & 53 & 2,433 & 1.85 & 0.04  & 0.08 \\
Oyente~\cite{luuMakingSmartContracts2016} & 3 & 83 & 769 & 3.49 & 0.39 & 0.70 \\
Securify~\cite{tsankovSecurifyPracticalSecurity2018a} & 1 & 3 & 1,004 & 25.00 & 0.10 & 0.19 \\
Semgrep~\cite{semgrepSemgrep2025} & 3,920 & 24,638  & 9,685  & 13.73 & 28.81 & 18.59 \\
Slither~\cite{feistSlitherStaticAnalysis2019} & 4,016 & 40,468  & 14,936  & 9.03 & 21.19 & 12.66 \\
Smartcheck~\cite{tikhomirovSmartCheckStaticAnalysis2018a} & 3,939  & 34,446   & 10,512   & 10.26  & 27.26  & 14.91  \\
Solhint~\cite{protofireSolhint2025} & 3,271   &  19,976 & 11,485   & 14.07 & 22.17 & 17.21 \\
\bottomrule
\end{tabular}
}
\vspace{-1em}
\label{tab:tool_comparison}
\end{table}

Our experiments reveal that \textbf{the highest F1-score achieved by any tool was only 18.59\%}, indicating a significant gap between current detection methods and the realities of real-world vulnerabilities. Notably, several tools, such as Manticore and Maian, failed to identify any true positives, which aligns with previous findings by Durieux \etal~\cite{durieuxEmpiricalReviewAutomated2020} and Sendner \etal~\cite{sendnerLargeScaleStudyVulnerability2024}.
This observation underscores the urgent need for more advanced detection tools and highlights the importance of utilizing large-scale, diverse datasets for accurate assessment and benchmarking.

Furthermore, \textbf{we observed a distinction between static analysis and symbolic execution approaches using our \tool dataset}. Static analysis tools, \eg, Semgrep, Slither, and Smartcheck, exhibited a higher number of true positives due to the lower environment requirements compared to symbolic execution tools like Oyente and Mythril. However, this advantage comes at the cost of a significantly higher false positive rate, leading to alarm fatigue that can hinder developers from addressing genuine vulnerabilities~\cite{dingVulnerabilityDetectionCode2024}. In contrast, symbolic execution tools, while more precise, suffered from timeouts on real-world complex contracts due to path explosion. These findings reveal gaps in existing detection approaches that might have been overlooked with smaller or less diverse datasets.

\begin{answerbox}
    \textbf{Answer to RQ4:} The \tool dataset demonstrates high practicality by evaluating existing security tools, revealing significant limitations in current detection capabilities.
\end{answerbox}

\section{Discussion}
\subsection{Implications}
Our new \tool dataset contains \files real-world vulnerabilities from \reports audit reports, which facilitates a comprehensive characterization of smart contract vulnerabilities in real-world DApp projects.

\begin{figure}[htbp]
    \centering
    \includegraphics[width=1.0\linewidth, clip]{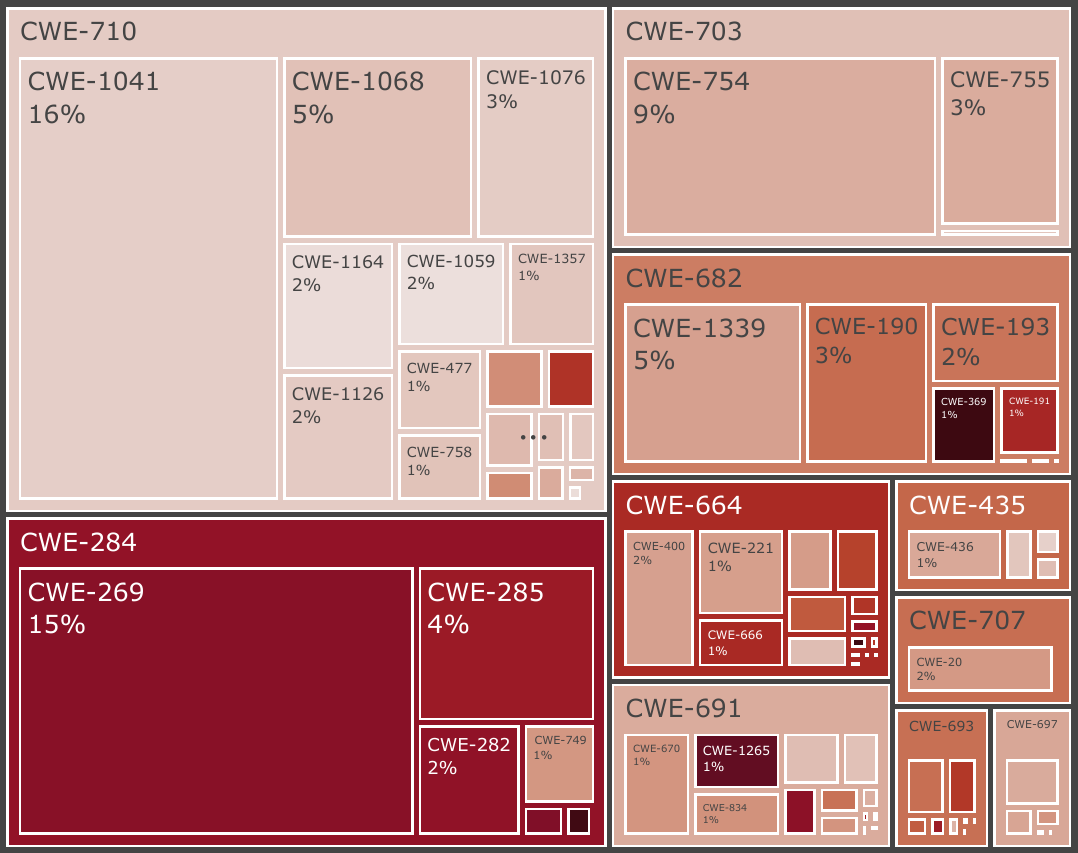}
    \vspace{-1.4em}
    \caption{Risk priority visualization of smart contract from CWE perspective. The size of each rectangle represents the frequency of occurrence, while the color depth indicates the severity level (darker colors represent higher severity). \protect\footnotemark}
    \label{fig:vuln-treemap}
    \vspace{-1.5em}
\end{figure}

\subsubsection{Vulnerability Frequency vs. Severity}

This study offers critical insights for security researchers and developers by examining the correlation between vulnerability frequency and severity, identifying high-impact vulnerabilities that warrant prioritized attention.

\textbf{Smart contracts with the same vulnerability type can exhibit different potential impacts.} For example, while smart contracts containing an \textit{Integer Bug} may lead to serious financial loss, the vulnerability could be insignificant if it appears only in functions that are never invoked.
To quantitatively represent the severity level of each vulnerability type, we follow the recommendations from previous work~\cite{scsvsSmartContractSecurity2021}, adopting the Common Vulnerability Scoring System v4.0 (\textit{CVSS v4.0})~\cite{nvdNVDVulnerabilityMetrics2025}, which is an open framework used to assess the severity of software vulnerabilities and provides a standardized way to rate them. 
We calculated the average \CVSS score $\bar{s}(c)$ for each CWE category present in our dataset as follows:

\begin{equation}
\bar{s}(c) = \frac{1}{|\mathcal{V}_c|} \sum_{v \in \mathcal{V}_c} s(v) 
\end{equation}

$\mathcal{V}_c$ denotes the set of vulnerability findings in our dataset that belong to the CWE category $c$, with each vulnerability denoted as $v \in \mathcal{V}_c$. $s(v)$ represents the \textit{CVSS} score of a vulnerability finding $v$. In our dataset, each vulnerability finding is assigned a severity level (i.e., \texttt{info}, \texttt{low}, \texttt{medium}, \texttt{high}, or \texttt{critical}) by auditors based on exploitation complexity and potential impacts. Each severity level corresponds to a \CVSS score according to the \CVSS standard.
$|\mathcal{V}_c|$ denotes the cardinality of the set $\mathcal{V}_c$, \ie, the total number of vulnerabilities in category $c$. This metric $\bar{s}(c)$ provides a quantitative representation of the average severity for each CWE category.

\footnotetext{A detailed interactive visualization is available in \href{https://FORGE-security.github.io}{\color{NavyBlue}{\url{https://FORGE-security.github.io}}}.}

To visualize the distribution, we created a treemap in Figure \ref{fig:vuln-treemap} that incorporates CWE hierarchical information, illustrating both the frequency and severity of smart contract vulnerabilities from a CWE perspective. As shown in Figure \ref{fig:vuln-treemap}, at the Pillar level, \textit{CWE-710: Improper Adherence to Coding Standards} emerges as the most frequent root cause. \textit{CWE-284: Improper Access Control} represents the vulnerability cause with the highest average severity. From a severity-frequency perspective, we observe that coding practice-related issues exhibit high frequency but relatively low severity. Conversely, there are distinct clusters of low-frequency but high-severity vulnerabilities, such as \textit{CWE-369: Divide By Zero}.

\begin{findingbox}
    \textbf{Finding 1:}
    High-severity smart contract vulnerabilities are not necessarily the most common.
\end{findingbox}

\subsubsection{Academic Research Priorities vs. Actual Security Concerns}

\begin{table}[htbp]
\centering
\caption{\tool average \CVSS score \textit{Top 10} vs. Academic Research Priority \textit{Top 10}}
\vspace{-0.8em}
\resizebox{1.0\columnwidth}{!}{
\begin{tabular}{cl||l}
\toprule[1.5pt]
\textbf{\#} & \textbf{CWE of \tool Top 10} & \makecell[l]{\textbf{Detection Count Top 10}\\\textbf{with CWE from Prior Research}} \\
\hline
\textbf{1} & \makecell[l]{CWE-940: Improper Verification of Source\\of a Communication Channel} & Reentrancy (28) CWE-1265 \\
\hline
\textbf{2} & CWE-369: Divide By Zero & Integer Bug (16) CWE-190/191 \\
\hline
\textbf{3} & \makecell[l]{CWE-347: Improper Verification of\\Cryptographic Signature} & \makecell[l]{Block-state Dependency (16) -\\CWE-829} \\ 
\hline
\textbf{4} & \makecell[l]{CWE-1265: Unintended Reentrant Invocation\\of Non-reentrant Code Via Nested Calls} & Control-flow Hijacking (15) CWE-691 \\
\hline
\textbf{5} & CWE-287: Improper Authentication & Mishandled Exception (15) CWE-703 \\
\hline
\textbf{6} & \makecell[l]{CWE-362: Concurrent Execution using\\Shared Resource with Improper\\Synchronization ('Race Condition')} & Assertion Failure (15) CWE-670 \\
\hline
\textbf{7} & CWE-269: Improper Privilege Management & Ether Leakage (14) CWE-282 \\
\hline
\textbf{8} & CWE-285: Improper Authorization & Suicidal Contract (14) CWE-749/826 \\
\hline
\textbf{9} & CWE-282: Improper Ownership Management & \makecell[l]{Transaction Origin Use (12)-\\CWE-477/284} \\
\hline
\textbf{10} & \makecell[l]{CWE-610: Externally Controlled Reference\\to a Resource in Another Sphere} & Freezing Ether (11) CWE-684 \\
\bottomrule[1.5pt]
\end{tabular}
}
\vspace{-0.5em}
\label{tab:blockchain_cwe}
\end{table}

The left half of table \ref{tab:blockchain_cwe} presents the top 10 specific CWE types in our dataset ranked by average \CVSS score, where higher scores indicate a greater need for security audits in practical applications.
To understand how existing research aligns with real-world security concerns, we compared these top 10 CWEs (called \tool Top 10) shown in Table~\ref{tab:blockchain_cwe} with those most frequently targeted vulnerabilities in current smart contract analysis tools. According to the comprehensive survey conducted by Zhang \etal~\cite{zhangDemystifyingExploitableBugs2023}, the right half of table \ref{tab:blockchain_cwe} presents the top 10 most frequently detected vulnerability types from 37 existing methods published on top-tier Software Engineering, Security, and Programming Language venues, which effectively representing the research community's primary focus.

\textbf{Our comparison reveals a significant misalignment between academic research priorities and real-world security concerns.} While existing tools heavily focus on well-known vulnerabilities such as reentrancy (28 tools), integer bugs (16 tools), and block-state dependency (16 tools), many of the most severe vulnerabilities identified in our dataset receive considerably less attention. For instance, \textit{CWE-940} and \textit{CWE-347}, which rank among the top 2 most severe vulnerabilities in our dataset, are rarely addressed by existing tools.
Many high-severity smart contract vulnerabilities related to financial business logic identified from our dataset require complex semantic understanding and context awareness, making them challenging targets for automated analysis. This explains why current tools tend to focus on more structurally identifiable issues. 

The misalignment highlights the importance of grounding security research in real-world data sources rather than focusing exclusively on machine-detectable vulnerabilities. Our findings suggest that the research community should recalibrate its focus to address the most severe security threats identified through actual security incidents and professional audits, even if these vulnerabilities present greater challenges for automated detection.

\begin{findingbox}
    \textbf{Finding 2:}
    There exists a significant gap between the vulnerabilities prioritized by academic research and those that pose the greatest risks in practice.
\end{findingbox}


\begin{figure}[htbp]
    \centering
    \includegraphics[width=0.85\linewidth]{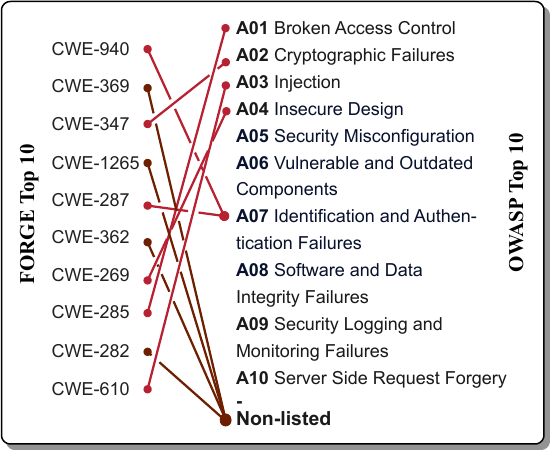}
    \vspace{-0.4em}
    \caption{\tool\textit{Top 10} vs. \textit{OWASP Top 10}~\protect\footnotemark}
    \vspace{-0.8em}
    \label{fig:slop-plot}
    
\end{figure}

\subsubsection{Smart Contract Vulnerabilities vs.  Traditional Software Vulnerabilities}
We conducted a comparative analysis between our smart contract vulnerability rankings and the widely recognized traditional software security risks. The \textit{OWASP Top 10} represents a broad consensus on the most critical security risks to web applications in the traditional software domain~\cite{owaspOWASPTopTen2024}. \MITRE provides a CWE perspective on the \textit{OWASP Top 10}, mapping each risk to one or more CWE root causes~\cite{mitreCWECWE1344Weaknesses2025}. To illustrate the divergence between the top 10 severity rankings from \tool and \OWASP root causes, we utilized a slope graph (Figure \ref{fig:slop-plot}) that visually demonstrates the disparity in vulnerability rankings.

\footnotetext{At the time of writing, the \textit{OWASP Top 10} 2021 was the most recent version available. The figure presents qualitative comparison results; the scale may be stretched for visualization purposes.}

\textbf{The comparison reveals significant differences between the vulnerability landscapes of smart contracts and traditional software applications.} Notably, while some \OWASP categories (A01-A04, A07) find corresponding instances in \tool\textit{Top 10}, they fail to encompass several of the most severe vulnerabilities in our dataset—including arithmetic issues, reentrancy problems, transaction order dependency (TOD), and ownership-related vulnerabilities that involve digital asset management and can lead to substantial losses if exploited. Additionally, A10 (Server-Side Request Forgery) does not have direct equivalents in our dataset.

This discrepancy between the \tool\textit{Top 10} and the \textit{OWASP Top 10} highlights the need for tailored security practices in the smart contract domain. The wealth of experience and best practices accumulated in traditional vulnerability management, such as prioritizing the most critical security issues for efficient resource allocation and establishing standardized security assessment criteria, cannot be directly transferred to the smart contract ecosystem. The unique characteristics of smart contracts and the blockchain environment demand the development of specialized security frameworks, tools, and practices that address the specific challenges and risks associated with decentralized applications. 

\begin{findingbox}
    \textbf{Finding 3:}
    Our analysis reveals a significant divergence between traditional software security and smart contract vulnerabilities, which highlights the need for tailored security practices in the smart contract domain.
\end{findingbox}

\subsubsection{For Ecosystem}
Our study offers several important implications for the smart contract ecosystem: \textit{(1) Research Focus Realignment.}
Our dataset provides empirical evidence to guide research efforts toward high-severity vulnerabilities that currently receive insufficient attention, helping bridge the gap between academic research and practical security needs.
\textit{(2) Security-First Development.}
The severity-ranked vulnerability distribution serves as a practical guideline for development practices, while our root cause analysis enables crucial feedback into the software development lifecycle for preventing entire vulnerability classes.
\textit{(3) Vulnerability Evolution Analysis.}
By applying \tool to audit reports across different time periods, researchers can track vulnerability pattern mutations, identify emerging threats, and understand the effectiveness of various security measures.
\textit{(4) Cross-Domain Knowledge Transfer.}
By CWE classification, our dataset enables comparing security challenges between traditional applications and smart contract systems, facilitating knowledge transfer across domains.

\subsection{Threats to Validity}

\paratitle{Internal Validity}
First, our dataset relies on LLM-based extraction and classification, which inherently introduces some imprecision when serving as ground truth datasets where high accuracy is crucial. This may impact practical utility. However, our evaluation in Section \ref{sec:rq2} demonstrates a high precision score for the generated entries, and our dataset is publicly accessible on GitHub, enabling community-driven validation and refinement through the issue tracking system.
Second, our approach utilizes the \CWE category, which was not originally designed for smart contract vulnerabilities. This potential mismatch could lead to inevitable inconsistencies in classification, as demonstrated in Section \ref{sec:rq3}. Nevertheless, our evaluation analysis reveals that existing vulnerabilities from real-world audit reports have already been classified effectively into \CWE, indicating sufficient coverage. Furthermore, our tree-of-thoughts method adaptively identifies the most appropriate vulnerability hierarchy level, mitigating potential inconsistencies.

\paratitle{External Validity}
First, our dataset exclusively comprises audit reports, potentially introducing selection bias. However, this approach actually strengthens dataset authenticity since these reports from 47 renowned teams represent popular, real-world projects subjected to a professional security evaluation, thereby enhancing ecological validity rather than diminishing it.
Second, our implementation employs LLama3 as the foundation model, raising concerns about how model performance might influence research conclusions, particularly with the emergence of new LLMs supporting longer context windows. This limitation is mitigated by our research objective, which focuses on dataset generation rather than context comprehension. Current LLMs, regardless of context length, still face fundamental challenges like ``lost in the middle'' effects. Within our framework, LLama3 demonstrates sufficient capability, and the modular design of \tool allows straightforward model substitution when more effective alternatives become available.

\section{Related work}
\paratitle{Smart Contract Vulnerability Datasets}
There has been substantial research effort in constructing high-quality smart contract vulnerability datasets. For example, Durieux \etal~\cite{durieuxEmpiricalReviewAutomated2020} introduced SmartBugs-wild, containing 47,398 solidity files from Ethereum with an average of only 204 lines of code per contract, and over 90\% using an outdated 0.4+ compiler version. ScrawlD~\cite{yashavantSujeetcScrawlD2021} incorporating vulnerability reports from multiple detection tools across 6,780 contracts. Both datasets were collected without manual vulnerability validation, including numerous false positives. Besides, their simplistic and outdated code samples fail to represent contemporary smart contract development practices. More recent efforts like Web3Bugs~\cite{zhangDemystifyingExploitableBugs2023} and DAppSCAN~\cite{zhengDAppSCANBuildingLargeScale2024} have attempted to provide higher-quality datasets by including deployed DApp audit reports. The larger one of these, DAppSCAN, covers 37 vulnerability types and includes 1,618 vulnerabilities.
While existing datasets rely on manual annotation processes that constrain their scope and evolution, \tool employs an automated approach, yielding a comprehensive dataset of \files Solidity files with an average of 2,577 lines of code per project, encompassing \vulnerabilities vulnerabilities across 296 CWE categories.

\paratitle{Vulnerability Classification}
Decentralized Application Security Project Top 10 (DASP10)~\cite{nccgroupDASPTOP102021}, Smart Contract Weakness Classification (SWC)~\cite{swcSmartContractWeakness2024} and Smart Contract Security Verification Standard (SCSVS)~\cite{scsvsSmartContractSecurity2021} have gained widespread adoption. However, these classifications often suffer from mixing different dimensions and providing non-orthogonal categories. Additionally, they have remained static for years, failing to evolve with the rapidly changing smart contract ecosystem~\cite{soudFlyOintmentEmpirical2023}.
In contrast, our work leverages the CWE classification, which provides a hierarchical, comprehensive, and regularly updated framework that encompasses both traditional software and smart contract vulnerability patterns.

\paratitle{Empirical Analysis of Real-World Smart Contract Vulnerabilities}
Recent studies have conducted large-scale analyses of the smart contract security ecosystem. Zhang et al.~\cite{zhangDemystifyingExploitableBugs2023} investigated 516 real-world vulnerabilities from 2021-2022, with a particular focus on machine-unauditable security bugs and their exploit patterns. Li et al.\cite{liStaticApplicationSecurity2024a} evaluated 8 Static Application Security Testing (SAST) tools using a newly created taxonomy and benchmark (98.85\% of vulnerabilities from previous work, 1.15\% from manual audits of real BNB projects), providing guidance on SAST tool evaluation and selection. Sendner et al.~\cite{sendnerLargeScaleStudyVulnerability2024} evaluated 18 vulnerability scanners across multiple datasets comprising over 4 million contracts, providing insights into the effectiveness of automated security tools. 
Compared to these works, our automated dataset construction approach enables more reliable and comprehensive empirical analysis through large-scale, real-world-sourced vulnerability findings. Moreover, from the unified perspective of the CWE, our work bridges the gap between smart contracts and traditional software, offering valuable comparative insights that were previously unavailable.

\section{Conclusion}
This paper presents \tool, an automated framework for constructing smart contract vulnerability datasets from real-world audit reports. Our approach employs a divide-and-conquer strategy and tree-of-thoughts reasoning to extract and classify vulnerabilities into CWE categories, addressing critical limitations in manual dataset construction. In 229.5 hours, \tool constructed a dataset containing \files Solidity files with \vulnerabilities vulnerabilities across 296 CWE categories, achieving expert-level classification consistency (\kalpha=0.87) and high extraction precision (\precision). Our empirical analysis reveals distinct vulnerability patterns in smart contracts compared to traditional software, significant misalignment between research focus and real-world priorities, and limited effectiveness of existing security tools (max 18.59\% F1). 
\tool advances smart contract security through automated real-world vulnerability dataset construction with a unified CWE classification, establishing essential foundations for both academic research and industrial practice.


\bibliographystyle{ACM-Reference-Format}
\bibliography{reference}

\end{document}